# On free vibration of piezoelectric nanospheres with surface effect

Bin Wu[1], Weiqiu Chen[1,2,3,4*], Chuanzeng Zhang[5]

[1] Department of Engineering Mechanics, Zhejiang University, Hangzhou 310027, China

[2] State Key Lab of CAD & CG, Zhejiang University, Hangzhou 310058, China

[3] Key Laboratory of Soft Machines and Smart Devices of Zhejiang Province, Zhejiang University, Hangzhou 310027, P.R. China

[4] Soft Matter Research Center, Zhejiang University, Hangzhou 310027, P.R. China

[5] Department of Civil Engineering, University of Siegen, Siegen D-57068, Germany

**Abstract:**

Surface effect responsible for some size-dependent characteristics can become distinctly important for piezoelectric nanomaterials with inherent large surface-to-volume ratio. In this paper, we investigate the surface effect on the free vibration behavior of a spherically isotropic piezoelectric nanosphere. Instead of directly using the well-known Huang-Yu surface piezoelectricity theory (HY theory), another general framework based on a thin shell layer model is proposed. A novel approach is developed to establish the surface piezoelectricity theory or the effective boundary conditions for piezoelectric nanospheres employing the state-space formalism. Three different sources of surface effect can be identified in the first-order surface piezoelectricity, i.e. the electroelastic effect, the inertia effect, and the thickness effect. It is found that the proposed theory becomes identical to the HY theory for a spherical material boundary if the transverse stress (TS) components are discarded and the electromechanical properties are properly defined. The nonaxisymmetric free vibration of a piezoelectric nanosphere with surface effect is then studied and an exact solution is obtained. In order to investigate the surface effect on the natural frequencies of piezoelectric nanospheres, numerical calculations are finally performed. Our numerical findings demonstrate that the surface effect, especially the thickness effect, may have a particularly significant influence on the free vibration of piezoelectric nanospheres. This work provides a more accurate prediction of the dynamic characteristics of piezoelectric nanospherical devices in Nano-Electro-Mechanical Systems (NEMS).

* Corresponding author. E-mail address: chenwq@zju.edu.cn

## 1. Introduction

With the rapid development of nanotechnology and advanced materials technology, piezoelectric nano-materials/structures have received more and more concerns due to their promising applications in future nanoelectromechanical systems (NEMS), such as nanosensors, nanoresonators, nanogenerators, nanoharvesters, diodes and field-effect transistors [1-3]. For example, the group led by Zhonglin Wang has utilized piezoelectric zinc-oxide (ZnO) nanowire arrays to demonstrate the first prototype nanogenerator [4-6], which can convert efficiently the nanoscale mechanical energy into electric energy. To date, these piezoelectric nano-materials/structures can be easily manufactured and synthesized in various configurations such as extremely thin films, nanoribbons, nanobeams, nanotubes, nanowires, nanospheres and nanoparticles [3,7,8]. For example, free-standing lead zirconate titanate (PZT) nanoparticles with a crystallite size between 10 and 15 nm were synthesized using the citrate nitrate (C/N) autocombustion method and their effective properties were measured using uniaxially pressed compacts at room temperature by Banerjee and Bose [9]. Faheem and Shoaib [10] described a modified sol-gel technique to synthesize nanocrystalline PZT powders and the obtained average particle size is less than 15 nm. Xu et al. [11] proposed a modified and controlled hydrothermal synthesis method to produce the nearly free-standing PZT nanoparticles in the size regime of about 4 nm. Zhang et al. [12] synthesized ZnO nanospheres with a typical wurtzite structure by a two-step self-assembly method and pointed out their applications in dye-sensitized solar cells. The controlled synthesis of cube- and sphere-like lithium niobate ($LiNbO_3$) nanoparticles via the single-source precursor was reported by Mohanty et al. [13]. Therefore, understanding the electromechanical behavior of these synthesized piezoelectric nano-materials/structures is particularly important both in the design and application stages, which in turn promotes advancements in the research of piezoelectricity and electromechanical theories at the nanoscale [1,3,7,14].

Due to the large ratio of surface area to volume, piezoelectric nanostructures exhibit various size-dependent characteristics which have been observed from either experiments or simulations [15-17]. Surface effect at the nanoscale, which takes account of the difference between the electromechanical properties of the bulk and its surface, is one of the reasonable and also successful explanations for this exotic characteristic. Physically, such material heterogeneity can be readily understood that atoms in or near a surface or interface have fewer bonding neighbors than atoms in the bulk counterpart [18,19]. Since controlled experiments on nanomaterials are extremely difficult and expensive while numerical simulations at the molecular and atomic levels are time-consuming, developing modified continuum mechanics models with the incorporation of surface effect is extremely important to the understanding and accurate prediction of static and dynamic behaviors of piezoelectric nano-materials/structures.

In order to analyze the surface effect in elastic or piezoelectric nanomaterials, there exist two principal approaches to establishing the theories of surface elasticity or piezoelectricity: one is the zero-thickness layer or Gibbs (geometrical) method wherein the surface layer is assumed as a mathematical surface with zero thickness; while the other is the finite-thickness layer method wherein the surface layer is regarded as a layer of extremely small but finite thickness [20]. For deformable non-piezoelectric elastic nanomaterials, Gurtin and Murdoch [21] presented a rigorous surface elasticity theory (hereafter referred to as the GM theory) based on the continuum mechanics framework, which treats the material surface as a two-dimensional (2D) heterogeneous membrane perfectly bonded to the underlying bulk material without slip. The presence of surface effect results in nonclassical boundary conditions on the bulk counterpart through the generalized Young-Laplace equations [22]. The GM theory has been utilized by many researchers to investigate the static and dynamic size-dependent behaviors and responses of nano-sized materials and structures [23-26]. In particular, the effect of surface stresses on vibration and buckling of piezoelectric nanowires was investigated by Wang and Feng [27] using the Euler-Bernoulli beam model, which takes account of the surface elasticity and the residual surface stresses, but ignores the surface piezoelectric effect.

As an extension of the GM theory, Huang and Yu [28] firstly established a surface piezoelectricity theory (abbreviated as the HY theory afterwards) in order to capture the surface effect on piezoelectric nanostructures. Based on the HY theory [28], which incorporates surface piezoelectric effect in addition to surface elasticity, residual surface stresses and electric displacements, substantial researches have been carried out to study the surface effect on static and dynamic electromechanical properties of various piezoelectric nanostructures [29-33]. Specifically, Yan and Jiang [29] utilized the Euler-Bernoulli beam theory to study the combined surface elasticity effect and surface electromechanical coupling on the vibration and buckling behaviors of piezoelectric nanobeams. The same authors also studied the surface effect on the electroelastic responses of a thin piezoelectric nanoplate using the conventional Kirchhoff plate model [30]. The surface effect on the propagation of the anti-plane shear waves and Rayleigh-Lamb-type waves in piezoelectric nanoplates was investigated by Zhang et al. [31] and Zhang et al. [32], respectively. In addition, Fang et al. [33] considered the effect of the interface energy on the size-dependent effective elastic constants of piezoelectric composites with spherically anisotropic nanoparticles under an external uniform strain. More recently, Chatzigeorgiou et al. [34] developed a phenomenological surface electrostatic theory in a consistent manner and investigated the surface effect on the overall response of nanoporous electric materials using the finite element method (FEM).

The zero-thickness 2D mathematical surface assumed in the theories described above including the GM and HY theories is an idealization of the actual situation because the real surface typically involves several atomic layers [15,19]. Following a systematic procedure for developing plate theories, Mindlin [35] was able to derive the effective boundary conditions on the plane boundary of a plate covered with a very thin layer, which were used later by Tiersten [36] in order to study the surface waves guided by thin films. It has been shown that the Mindlin-Tiersten model is exactly the same as the linearized GM theory if the residual surface tension is absent and the elastic properties are properly defined [19,37,38]. Furthermore, Tiersten et al. [39] obtained the governing equations for an anisotropic surface membrane with residual surface tension from the three-dimensional (3D) equations by taking only the three stress components acting in the plane of the membrane to be nonzero, assuming the uniformity of all membrane variables across the thickness, and integrating with respect to the thickness. The resulting equations for the surface membrane obtained by Tiersten et al. [39] are the same as those formulated by Gurtin and Murdoch [21]. Mindlin-Tiersten's treatment was followed by some other researchers, who however adopted the simple Taylor's series expansion to derive the effective boundary conditions of an elastic or a piezoelectric surface/interface material layer [40-43]. Theoretically, the effective boundary conditions accurate up to an arbitrary order can be obtained by truncating the series expansion. In particular, Bövik [41] made a comparison between the Mindlin-Tiersten model and the first-order effective boundary conditions for elastic surface waves guided by thin layers and showed that the latter gives a much better prediction than the former for the Rayleigh-type surface waves.

However, the mathematical manipulation based on Taylor's series expansion will become extremely tedious and cumbersome when material anisotropy and multi-field coupling are involved. Recently, based on Mindlin and Tiersten's thin layer model [35,36], a novel approach was proposed by Chen [38,44], which can be easily applied to establish a surface piezoelectricity theory based on the state-space formalism. Using the method developed by Chen [38,44], surface effects on wave propagation in anisotropic nanocylinders [19], piezoelectric half-spaces [44], piezoelectric nanoplates [38], piezoelectric nanocylinders [45], and multiferroic nanoplates [46], have been systemically investigated.

The goal of this paper is to extend our previous works for plane boundary [38] and cylindrical boundary [19] to spherical boundary, and investigate the free vibration behavior of a spherically isotropic piezoelectric nanosphere. We first give the 3D basic governing equations and the corresponding state-space formalism in spherical coordinates in Section 2. By means of the state-space formalism, a surface piezoelectricity theory for a spherical material boundary is established in Section 3. We further compare the proposed surface piezoelectric theory with the HY theory in Section 4. The frequency equations for the free vibration of spherically isotropic piezoelectric nanospheres with surface effect are then obtained analytically in Section 5. Numerical results are finally presented to elucidate the surface effect on the natural frequencies of a PZT-5H nanosphere in Section 6. We draw some conclusions in Section 7 to summarize our main findings. Some related mathematical expressions and derivations are provided in Appendices A and B.

**2. Basic equations and state-space formalism in spherical coordinates**

Consider a spherically isotropic piezoelectric nanosphere with a radial polarization. The spherical coordinates $(r, \theta, \varphi)$ are used with the origin located at the center of the material anisotropy. At the nanoscale, atoms in or near the surfaces have fewer bonding neighbors than atoms in the bulk counterpart. As a consequence, the electromechanical properties of the surface are, in general, somewhat different from those of the bulk [46]. In addition, the surface region of nano-sized structures, a transition zone between its homogeneous bulk and the vacuum, typically penetrates a few atomic layers into the bulk material, and thus has a certain thickness [15]. Accordingly, the nanosphere can be modeled as a core-shell structure composed of a spherical core of radius $r_0$ and a thin spherical shell surface layer of thickness $h = r_1 - r_0$, with $r_1$ being the outer radius of the shell. The spherical shell surface layer and the spherical core are concentric but have different electromechanical properties.

For spherically isotropic piezoelectric materials with a radial polarization, the linear constitutive relations are given by [47]

$$\begin{aligned}
&\sigma_{\theta\theta} = c_{11}\varepsilon_{\theta\theta} + c_{12}\varepsilon_{\varphi\varphi} + c_{13}\varepsilon_{rr} - e_{31}E_r, \quad \sigma_{\varphi\varphi} = c_{12}\varepsilon_{\theta\theta} + c_{11}\varepsilon_{\varphi\varphi} + c_{13}\varepsilon_{rr} - e_{31}E_r, \\
&\sigma_{rr} = c_{13}\varepsilon_{\theta\theta} + c_{13}\varepsilon_{\varphi\varphi} + c_{33}\varepsilon_{rr} - e_{33}E_r, \quad \sigma_{r\varphi} = 2c_{44}\varepsilon_{r\varphi} - e_{15}E_\varphi, \quad \sigma_{\theta\varphi} = 2c_{66}\varepsilon_{\theta\varphi}, \\
&\sigma_{r\theta} = 2c_{44}\varepsilon_{r\theta} - e_{15}E_\theta, \quad D_\theta = 2e_{15}\varepsilon_{r\theta} + \kappa_{11}E_\theta, \quad D_\varphi = 2e_{15}\varepsilon_{r\varphi} + \kappa_{11}E_\varphi, \\
&D_r = e_{31}\left(\varepsilon_{\theta\theta} + \varepsilon_{\varphi\varphi}\right) + e_{33}\varepsilon_{rr} + \kappa_{33}E_r
\end{aligned} \quad (1)$$

where $\sigma_{ij}$ and $\varepsilon_{ij}$ are the stress and strain tensors, respectively; $D_i$ and $E_i$ are the electric displacement and electric field vectors, respectively; $c_{ij}$, $e_{ij}$ and $\kappa_{ij}$ are the elastic, piezoelectric, and dielectric constants, respectively.

The strain-displacement and electric field-potential relations in spherical coordinates are

$$\varepsilon_{rr} = \frac{\partial u_r}{\partial r}, \quad \varepsilon_{\theta\theta} = \frac{1}{r}\frac{\partial u_\theta}{\partial \theta} + \frac{u_r}{r}, \quad \varepsilon_{\varphi\varphi} = \frac{1}{r\sin\theta}\frac{\partial u_\varphi}{\partial \varphi} + \frac{u_r}{r} + \frac{u_\theta}{r}\cot\theta,$$

$$2\varepsilon_{\theta\varphi} = \frac{1}{r}(\frac{\partial u_\varphi}{\partial \theta} - u_\varphi \cot\theta) + \frac{1}{r\sin\theta}\frac{\partial u_\theta}{\partial \varphi}, \quad 2\varepsilon_{r\varphi} = \frac{1}{r\sin\theta}\frac{\partial u_r}{\partial \varphi} + \frac{\partial u_\varphi}{\partial r} - \frac{u_\varphi}{r}, \quad (2)$$

$$2\varepsilon_{r\theta} = \frac{1}{r}\frac{\partial u_r}{\partial \theta} + \frac{\partial u_\theta}{\partial r} - \frac{u_\theta}{r}, \quad E_r = -\frac{\partial \Phi}{\partial r}, \quad E_\theta = -\frac{1}{r}\frac{\partial \Phi}{\partial \theta}, \quad E_\varphi = -\frac{1}{r\sin\theta}\frac{\partial \Phi}{\partial \varphi}$$

where $u_i$ and $\Phi$ are the mechanical displacement vector and the electric potential, respectively.

In spherical coordinates, the differential equations of motion without body forces and the Gauss's law for electrostatics in the absence of free charges can be written as

$$\frac{\partial \sigma_{rr}}{\partial r} + \frac{1}{r}\frac{\partial \sigma_{r\theta}}{\partial \theta} + \frac{1}{r\sin\theta}\frac{\partial \sigma_{r\varphi}}{\partial \varphi} + \frac{1}{r}(2\sigma_{rr} - \sigma_{\theta\theta} - \sigma_{\varphi\varphi} + \sigma_{r\theta}\cot\theta) = \rho \ddot{u}_r,$$

$$\frac{\partial \sigma_{r\theta}}{\partial r} + \frac{1}{r}\frac{\partial \sigma_{\theta\theta}}{\partial \theta} + \frac{1}{r\sin\theta}\frac{\partial \sigma_{\theta\varphi}}{\partial \varphi} + \frac{1}{r}[(\sigma_{\theta\theta} - \sigma_{\varphi\varphi})\cot\theta + 3\sigma_{r\theta}] = \rho \ddot{u}_\theta,$$

$$\frac{\partial \sigma_{r\varphi}}{\partial r} + \frac{1}{r}\frac{\partial \sigma_{\theta\varphi}}{\partial \theta} + \frac{1}{r\sin\theta}\frac{\partial \sigma_{\varphi\varphi}}{\partial \varphi} + \frac{1}{r}(2\sigma_{\theta\varphi}\cot\theta + 3\sigma_{r\varphi}) = \rho \ddot{u}_\varphi, \quad (3)$$

$$\frac{1}{r^2}\frac{\partial}{\partial r}(r^2 D_r) + \frac{1}{r\sin\theta}\frac{\partial}{\partial \theta}(D_\theta \sin\theta) + \frac{1}{r\sin\theta}\frac{\partial D_\varphi}{\partial \varphi} = 0$$

where $\rho$ is the mass density and the superimposed dot denotes derivative with respect to time $t$.

In this paper, the state-space formalism will be utilized to derive the surface piezoelectricity theory for the spherical material boundary (i.e., the spherical shell surface layer). The state-space method has several particular advantages over the displacement-based method in solving many practical problems, and the interested reader is referred to Ding and Chen [47] and Chen and Ding [48] as well as the references cited therein for more details.

If we choose $\mathbf{Y}_1 = [u_r, u_\theta, u_\varphi, \Phi]^T$ and $\mathbf{Y}_2 = [\sigma_{rr}, \sigma_{r\theta}, \sigma_{r\varphi}, D_r]^T$ (the superscript T signifies transpose) as the generalized displacement vector and the generalized stress vector, respectively, and combine them into a state vector, the state equation can be readily derived from Eqs. (1)-(3) following a standard way [47,48]. For simplicity, we directly give the state equation without the detailed derivations as follows

$$\frac{\partial}{\partial r}\begin{Bmatrix}\mathbf{Y}_1\\\mathbf{Y}_2\end{Bmatrix} = \mathbf{M}(c_{ij}, e_{ij}, \kappa_{ij}; r, \theta; \partial_\theta, \partial_\varphi, \partial_t)\begin{Bmatrix}\mathbf{Y}_1\\\mathbf{Y}_2\end{Bmatrix} = \begin{bmatrix}\mathbf{M}_{11} & \mathbf{M}_{12}\\\mathbf{M}_{21} & \mathbf{M}_{22}\end{bmatrix}\begin{Bmatrix}\mathbf{Y}_1\\\mathbf{Y}_2\end{Bmatrix} \qquad (4)$$

where $\partial_\eta = \partial/\partial_\eta$ with $\eta = \theta, \varphi, t$; and $\mathbf{M}$ is the $8 \times 8$ system matrix, with its four partitioned $4 \times 4$ sub-matrices $\mathbf{M}_{ij}$ being given in Appendix A.

## 3. Surface piezoelectricity theory for a spherical material boundary

As described above, the electromechanical properties of the surface region for a piezoelectric nanosphere are generally distinct from those of the bulk material. In this section, the method proposed by Chen [19,38] is exploited to establish the surface piezoelectricity for a spherical material boundary of a piezoelectric nanosphere without residual stress and electric displacement. The starting point is the state equation (4). The spherical material boundary is seen as a thin piezoelectric spherical shell with thickness $h$. Since the spherical shell surface layer is extremely thin, the so-called effective boundary conditions which govern the motion of the thin shell can be obtained based on the state-space formalism in an approximate way, instead of directly treating the surface spherical shell as a different material phase.

For clarity, a superscript $s$ will be adopted to indicate the quantities that are associated with the surface spherical shell. Applying Eq. (4) to the thin spherical shell $r_0 \leq r \leq r_1$, we get

$$\frac{\partial}{\partial r}\begin{Bmatrix} \mathbf{Y}_1^s \\ \mathbf{Y}_2^s \end{Bmatrix} = \mathbf{M}^s(c_{ij}^s, e_{ij}^s, \kappa_{ij}^s; r, \theta; \partial_\theta, \partial_\varphi, \partial_t)\begin{Bmatrix} \mathbf{Y}_1^s \\ \mathbf{Y}_2^s \end{Bmatrix} \tag{5}$$

From Eq. (5), it can be seen that the system matrix $\mathbf{M}^s$ is not only related to the three partial differential operators $\partial_\theta, \partial_\varphi, \partial_t$, but also depend on the radial coordinate $r$. However, considering the fact that the surface spherical shell is very thin (i.e. $h$ is very small), and as an approximation, we may take $r \approx r_0$ to obtain

$$\frac{\partial}{\partial r}\begin{Bmatrix} \mathbf{Y}_1^s \\ \mathbf{Y}_2^s \end{Bmatrix} = \mathbf{M}_0^s(r_0; \partial_\theta, \partial_\varphi, \partial_t)\begin{Bmatrix} \mathbf{Y}_1^s \\ \mathbf{Y}_2^s \end{Bmatrix} \tag{6}$$

where $\mathbf{M}_0^s = \mathbf{M}^s|_{r=r_0}$. By treating the partial differential operators in $\mathbf{M}_0^s$ as usual parameters, the solution to Eq. (6) can be formally written as

$$\begin{Bmatrix} \mathbf{Y}_1^s(r) \\ \mathbf{Y}_2^s(r) \end{Bmatrix} = \exp[\mathbf{M}_0^s(r - r_0)]\begin{Bmatrix} \mathbf{Y}_1^s(r_0) \\ \mathbf{Y}_2^s(r_0) \end{Bmatrix} \tag{7}$$

Setting $r = r_1$ in Eq. (7) leads to the following relation between the state vectors at the outer and inner surfaces of the thin shell

$$\begin{Bmatrix} \mathbf{Y}_1^s(r_1) \\ \mathbf{Y}_2^s(r_1) \end{Bmatrix} = \exp(\mathbf{M}_0^s h)\begin{Bmatrix} \mathbf{Y}_1^s(r_0) \\ \mathbf{Y}_2^s(r_0) \end{Bmatrix} \equiv \mathbf{F}\begin{Bmatrix} \mathbf{Y}_1^s(r_0) \\ \mathbf{Y}_2^s(r_0) \end{Bmatrix} \tag{8}$$

where $\mathbf{F} = \exp(\mathbf{M}_0^s h)$ is the transfer matrix. According to the definition of the matrix exponential, the following Taylor's series expansion can be made

$$\mathbf{F} = \mathbf{I} + \mathbf{M}_0^s h + \frac{1}{2}(\mathbf{M}_0^s)^2 h^2 + \cdots + \frac{1}{n!}(\mathbf{M}_0^s)^n h^n + O(h^{n+1}) \tag{9}$$

From Eqs. (8) and (9), we can obtain the following relations, which are accurate up to

the first order, as

$$\begin{Bmatrix} \mathbf{Y}_1^s(r_1) \\ \mathbf{Y}_2^s(r_1) \end{Bmatrix} = (\mathbf{I} + \mathbf{M}_0^s h) \begin{Bmatrix} \mathbf{Y}_1^s(r_0) \\ \mathbf{Y}_2^s(r_0) \end{Bmatrix} \tag{10}$$

For the free vibration problem, the outer surface $r = r_1$ of the thin spherical shell is free of tractions as well as free surface charges, i.e.,

$$\mathbf{Y}_2^s(r_1) = \mathbf{0} \tag{11}$$

In addition, the state variables at the inner surface of the spherical shell should be equal to those of the spherical core material at $r = r_0$, i.e.,

$$\mathbf{Y}_1^s(r_0) = \mathbf{Y}_1(r_0), \quad \mathbf{Y}_2^s(r_0) = \mathbf{Y}_2(r_0) \tag{12}$$

where $\mathbf{Y}_1$ and $\mathbf{Y}_2$ without the superscript $s$ denote the state vectors associated with the spherical core material. Substitution of Eqs. (11) and (12) into the last three equations in Eq. (10) yields the first-order effective boundary conditions for the surface spherical shell as

$$\mathbf{Y}_2(r_0) + [\mathbf{M}_{021}^s \mathbf{Y}_1(r_0) + \mathbf{M}_{022}^s \mathbf{Y}_2(r_0)]h = \mathbf{0} \tag{13}$$

where $\mathbf{M}_{0ij}^s$ are the $4 \times 4$ partitioned sub-matrices of the matrix $\mathbf{M}_0^s$. Equation (13) is the first-order effective boundary conditions or the first-order surface piezoelectricity theory for a spherical material boundary in the absence of residual stresses and electric displacements. Just like the cases of isotropic nanoplates [38] and nanocylinders [19], three different sources of surface effect can be identified in this theory, i.e. the electroelastic effect, the inertia effect and the thickness effect. The thickness effect, which corresponds to the last term in Eq. (13), is associated with the three transverse stress (TS) components $(\sigma_{rr}, \sigma_{r\theta}, \sigma_{r\varphi})$ and the radial electric displacement $D_r$ in the interior of the surface spherical shell, the inertia effect corresponds to the second-order time-derivatives in the second term, while the rest in the second term is attributed to the electroelasticity of the piezoelectric material surface.

Note that if the surface effect is discarded, the first-order effective boundary conditions (13) degenerate to the classical homogeneous boundary conditions. In addition, effective boundary conditions of an arbitrary order governing the surface piezoelectric spherical shell may also be readily obtained by appropriately truncating the Taylor's series in Eq. (9). Furthermore, if different boundary conditions are imposed on the outer surface of the thin spherical shell, we can deduce the corresponding effective boundary conditions in a similar manner [46]. The first-order effective boundary conditions will be utilized in Section 5 to investigate the surface effect on the free vibration of piezoelectric nanospheres.

In the above derivations, we have assumed that all the material constants are homogeneous over any spherical surface, and hence the resulting surface material properties are also homogeneous over the surface. If the inhomogeneous case should be considered, the state equation (4) should be also modified accordingly.

In addition, any dependence of the material properties on the curvature has been ignored in our analysis. Thus, the surface material parameters are constant for a given piezoelectric nanomaterial and independent of the radius of the nanosphere. It is noted here that the curvature-dependent surface and interface elasticity theories [49-52], which generalize the GM theory to take into account the effect of the flexural resistance of the elastic films, have been also proposed previously for elastic nanostructures. Indeed, the development of a unified surface/interface piezoelectricity theory with the curvature-dependence for piezoelectric nanomaterials is very interesting, but will be out of the scope of the present study.

## 4. Comparison with the HY theory

In the absence of residual stresses and electric displacements, the effective boundary conditions of the HY theory for a spherically isotropic material surface in spherical coordinates are given in Appendix B. If we neglect the three TS components but retain the radial electric displacement in the last term of Eq. (13), the first-order surface piezoelectricity theory (13) becomes

$$[\sigma_{rr}, \sigma_{r\theta}, \sigma_{r\varphi}, D_r]^\mathrm{T} + \mathbf{M}^s_{021} h[u_r, u_\theta, u_\varphi, \Phi]^\mathrm{T} + \mathbf{M}^s_{022} h[0,0,0,D_r]^\mathrm{T} = \mathbf{0}, \quad (r = r_0) \quad (14)$$

We can refer to the effect, which takes account of the three TS components, as the TS effect. In view of the formulae in Appendix A, we can obtain from Eq. (14)

$$\sigma_{rr} + \left[\rho^s h \frac{\partial^2}{\partial t^2} + \frac{2(k_1^s + k_2^s)h}{r_0^2}\right] u_r + \frac{(k_1^s + k_2^s)h}{r_0^2}\left(\cot\theta + \frac{\partial}{\partial\theta}\right) u_\theta$$

$$+ \frac{(k_1^s + k_2^s)h}{r_0^2 \sin\theta} \frac{\partial u_\varphi}{\partial\varphi} - \frac{2\gamma^s h}{\alpha^s r_0} D_r = 0,$$

$$\sigma_{r\theta} - \frac{(k_1^s + k_2^s)h}{r_0^2} \frac{\partial u_r}{\partial\theta} + \left[\rho^s h \frac{\partial^2}{\partial t^2} - \frac{h}{r_0^2}\left(k_1^s \nabla_4^2 - k_2^s + \frac{c_{66}^s}{\sin^2\theta}\frac{\partial^2}{\partial\varphi^2}\right)\right] u_\theta$$

$$- \frac{h}{r_0^2 \sin\theta}\left[(k_2^s + c_{66}^s)\frac{\partial^2}{\partial\varphi\partial\theta} - (k_1^s + c_{66}^s)\cot\theta\frac{\partial}{\partial\varphi}\right] u_\varphi + \frac{\gamma^s h}{\alpha^s r_0}\frac{\partial D_r}{\partial\theta} = 0,$$

$$\sigma_{r\varphi} - \frac{(k_1^s + k_2^s)h}{r_0^2 \sin\theta}\frac{\partial u_r}{\partial\varphi} - \frac{h}{r_0^2 \sin\theta}\left[(k_2^s + c_{66}^s)\frac{\partial^2}{\partial\theta\partial\varphi} + (k_1^s + c_{66}^s)\cot\theta\frac{\partial}{\partial\varphi}\right] u_\theta$$

$$+ \left\{\rho^s h \frac{\partial^2}{\partial t^2} - \frac{h}{r_0^2}\left[\frac{k_1^s}{\sin^2\theta}\frac{\partial^2}{\partial\varphi^2} + c_{66}^s(\nabla_4^2 + 1)\right]\right\} u_\varphi + \frac{\gamma^s h}{\alpha^s r_0 \sin\theta}\frac{\partial D_r}{\partial\varphi} = 0,$$

$$D_r + \frac{k_3^s h}{r_0^2}\nabla_1^2 \Phi - \frac{2h}{r_0} D_r = 0 \qquad (15)$$

Now establish the following relations between the surface material constants in our theory and those in the HY theory:

$$\bar{\rho}_0^Y = \rho^s h, \ \bar{k}_1^Y = k_1^s h, \ \bar{k}_2^Y = k_2^s h, \ \bar{c}_{66}^Y = c_{66}^s h, \ \bar{\kappa}_{11}^Y = k_3^s h, \ \frac{\bar{e}_{31}^Y}{\bar{\kappa}_{33}^Y} = \frac{\gamma^s}{\alpha^s}, \ D_r^Y = hD_r\big|_{r=r_0} \quad (16)$$

where $\bar{\rho}_0^Y$, $\bar{k}_1^Y$, etc. are the surface material constants defined in the HY theory. It is easily found that Eq. (15) is the same as Eq. (B7) in Appendix B, which corresponds to the HY theory when the surface residual stresses and electric displacements are absent. Thus, we can conclude that, the proposed surface piezoelectricity theory based on the state-space formalism can lead to the same equations as the HY theory, provided that the TS effect in the first-order approximation is omitted and the surface material properties are properly defined according to Eq. (16). Interestingly, it can be seen from Eq. (16) that the surface material constants $\bar{\rho}_0^Y$, $\bar{k}_1^Y$, etc. are merely the scaled versions of their bulk counterparts. This fact has also been noticed by Gurtin and Murdoch for a plane material surface [37] and by Chen et al. for a cylindrical surface [19].

It is noted that, via the thin layer model, the surface radial electric displacement $D_r^Y$ in the 2D HY theory appears to be the bulk counterpart multiplied by the thickness of the surface layer, as seen from the last one of Eq. (16). However, the physical reality of this surface electric displacement for a zero-thickness mathematical surface still needs to be explored.

## 5. Free vibration of piezoelectric nanospheres with surface effect

### 5.1. *Free vibration solution for the spherical core*

Consider the general nonaxisymmetric free vibration of a spherically isotropic piezoelectric nanosphere with the radial polarization. In spherical coordinates, three displacement functions $w$, $\psi$ and $G$ are introduced as follows:

$$u_r = w, \quad u_\theta = -\frac{1}{\sin\theta}\frac{\partial \psi}{\partial \varphi} - \frac{\partial G}{\partial \theta}, \quad u_\varphi = \frac{\partial \psi}{\partial \theta} - \frac{1}{\sin\theta}\frac{\partial G}{\partial \varphi} \tag{17}$$

Substituting Eq. (17) into Eq. (2) and then into Eqs. (1) and (3), we obtain the governing equations of the spherical core in terms of $\psi$, $G$, $w$ and $\Phi$, which can be solved by assuming

$$[\psi, G, w, \Phi] = r_0 \left[ U_n(\xi), V_n(\xi), W_n(\xi), (e_{33}/\varepsilon_{33}) X_n(\xi) \right] S_n^m(\theta, \varphi) \exp(i\omega t) \tag{18}$$

where $i = \sqrt{-1}$ is the imaginary unit; $S_n^m(\theta,\varphi) = P_n^m(\cos\theta)\exp(im\varphi)$ are spherical harmonics and $P_n^m(\cos\theta)$ are the associated Legendre polynomials; $n$ and $m$ are integers; $\omega$ is the circular frequency; and $\xi = r/r_0$ is the dimensionless radial coordinate. The four dimensionless unknown functions $U_n(\xi)$, $V_n(\xi)$, $W_n(\xi)$ and $X_n(\xi)$ in Eq. (18) satisfy

$$\begin{aligned}
&\xi^2 U_n'' + 2\xi U_n' + \left\{\Omega^2\xi^2 - \left[2 + (n^2+n-2)(f_1-f_2)/2\right]\right\} U_n = 0, \\
&\xi^2 W_n'' + 2\xi W_n' + \left(\Omega^2\xi^2/f_4 + p_1\right)W_n - p_2\xi V_n' - p_3 V_n + q_1\xi^2 X_n'' + q_2\xi X_n' + q_3 X_n = 0, \\
&\xi^2 V_n'' + 2\xi V_n' + \left(\Omega^2\xi^2 + p_4\right)V_n - p_5\xi W_n' - p_6 W_n + q_4\xi X_n' + q_5 X_n = 0, \\
&\xi^2 X_n'' + 2\xi X_n' + q_6 X_n - \xi^2 W_n'' - p_7\xi W_n' - p_8 W_n - p_9\xi V_n' - p_{10} V_n = 0
\end{aligned} \tag{19}$$

where $\Omega = \omega r_0 \sqrt{\rho/c_{44}}$ is the dimensionless frequency; a prime denotes the differentiation with respect to $\xi$; $p_i$, $q_i$ and $f_i$ are the dimensionless material constants defined in Ding and Chen [47], which are omitted here for brevity.

It can be seen from Eq. (19) that the unknown $U_n$ is uncoupled from the other three unknowns $V_n$, $W_n$ and $X_n$. In fact, the first one of Eq. (19) is a second-order uncoupled partial differential equation in $U_n$ while the last three of Eq. (19) form a set of coupled partial differential equations in $V_n$, $W_n$ and $X_n$. The solutions to these equations for a piezoelectric hollow sphere could be readily obtained [47,53]. The solutions for an inhomogeneous piezoelectric hollow sphere were also derived by Chen [54]. Therefore, we will directly present the final results by omitting the mathematical details. However, it should be emphasized here that for a piezoelectric solid sphere, we can only retain those solutions such that the mechanical displacements and the electric potential remain bounded at the origin $r=0$ (i.e., the center of the sphere).

The solution to the first one of Eq. (19) can be written as

$$U_n(\xi) = \xi^{-1/2} B_{n1} J_\eta(\Omega\xi), \quad (n \geq 1) \tag{20}$$

where $\eta^2 = \left[9 + 2(n^2 + n - 2)(f_1 - f_2)\right]/4 > 0$, $J_\eta$ is the Bessel function of the first kind, and $B_{n1}$ is an arbitrary constant.

For $n \geq 1$, we note that $\xi = 0$ is a regular singular point of the coupled system in the last three formulae of Eq. (19). The matrix Frobenius power series method developed by Ding et al. [55] could be used to obtain the following general solution:

$$W_n(\xi) = \sum_{j=1}^{3} C_{nj} W_{nj}(\xi), \quad V_n(\xi) = \sum_{j=1}^{3} C_{nj} V_{nj}(\xi), \quad X_n(\xi) = \sum_{j=1}^{3} C_{nj} X_{nj}(\xi), \quad (n \geq 1) \tag{21}$$

where $C_{nj}$ are arbitrary constants; $W_{nj}$, $V_{nj}$ and $X_{nj}$ are convergent and infinite series in the variable $\xi$ [47,55].

The mode $n=0$ corresponding to the purely radial vibration is a special case, in which the function $V_n(\xi)$ contributes nothing to the electroelastic field and only the two unknowns $W_n$ and $X_n$ are coupled together. The corresponding solutions are

$$W_0(\xi) = \sum_{j=1}^{2} C_{0j} W_{0j}(\xi), \quad X_0(\xi) = \sum_{j=1}^{2} C_{0j} X_{0j}(\xi), \quad (n=0) \tag{22}$$

With the above solutions, the field variables in the spherical core of the piezoelectric nanosphere can be expressed in dimensionless functions $U_n$, $V_n$, $W_n$ and $X_n$ as [47,53]

$$
\begin{aligned}
&u_r = r_0 W_n(\xi) S_n^m(\theta,\varphi), \quad u_\theta = -r_0 \left[ \frac{U_n(\xi)}{\sin\theta} \frac{\partial S_n^m(\theta,\varphi)}{\partial \varphi} + V_n(\xi) \frac{\partial S_n^m(\theta,\varphi)}{\partial \theta} \right], \\
&u_\varphi = r_0 \left[ U_n(\xi) \frac{\partial S_n^m(\theta,\varphi)}{\partial \theta} - \frac{V_n(\xi)}{\sin\theta} \frac{\partial S_n^m(\theta,\varphi)}{\partial \varphi} \right], \quad \Phi = \frac{e_{33}}{\varepsilon_{33}} r_0 X_n(\xi) S_n^m(\theta,\varphi), \\
&\sigma_{rr} = c_{13} R_n(\xi) S_n^m(\theta,\varphi), \quad \sigma_{r\theta} = c_{44} \left[ T_n(\xi) \frac{\partial S_n^m(\theta,\varphi)}{\partial \theta} - Y_n(\xi) \frac{1}{\sin\theta} \frac{\partial S_n^m(\theta,\varphi)}{\partial \varphi} \right], \\
&\sigma_{r\varphi} = c_{44} \left[ T_n(\xi) \frac{1}{\sin\theta} \frac{\partial S_n^m(\theta,\varphi)}{\partial \varphi} + Y_n(\xi) \frac{\partial S_n^m(\theta,\varphi)}{\partial \theta} \right], \quad D_r = e_{33} Z_n(\xi) S_n^m(\theta,\varphi)
\end{aligned}
\tag{23}
$$

where the common time-harmonic factor $\exp(i\omega t)$ appearing in all field variables is omitted; $R_n(\xi)$, $T_n(\xi)$, $Y_n(\xi)$ and $Z_n(\xi)$ are defined as

$$
\begin{aligned}
Y_n(\xi) &= U'_n - U_n/\xi, \\
R_n(\xi) &= 2W_n/\xi + V_n n(n+1)/\xi + (f_4/f_3)W'_n + (f_8/f_3)X'_n, \\
T_n(\xi) &= W_n/\xi - V'_n + V_n/\xi + f_5 f_8 X_n/\xi, \\
Z_n(\xi) &= 2f_6 W_n/\xi + f_6 V_n n(n+1)/\xi + W'_n - X'_n
\end{aligned}
\tag{24}
$$

### 5.2. *Effective boundary conditions*

Although the surface spherical shell of the piezoelectric nanosphere is governed by the first-order effective boundary conditions (13), in the following, we will use Eq. (14) or (15) instead of Eq. (13) to derive, just for illustration, the frequency equation for the piezoelectric nanosphere with surface effect. This is because the associated formulae become simpler, while the derivation process remains the same when the TS effect is taken into account. Substituting Eq. (23) into the first and fourth formulae of

Eq. (15), we obtain

$$R_n + \frac{h}{r_0}\left\{\frac{k_1^s + k_2^s}{c_{13}}\left[2W_n + n(n+1)V_n\right] - \frac{\rho^s\Omega^2 c_{44}}{\rho c_{13}}W_n - \frac{2e_{33}\gamma^s}{c_{13}\alpha^s}Z_n\right\} = 0, \quad (\xi = 1,\ n \geq 0)$$

$$\left(1 - \frac{2h}{r_0}\right)Z_n - \frac{h}{r_0}\frac{k_3^s}{\varepsilon_{33}}n(n+1)X_n = 0, \quad (\xi = 1,\ n \geq 0)$$
(25)

where the identity $\nabla_1^2 S_n^m(\theta,\varphi) + n(n+1)S_n^m(\theta,\varphi) = 0$ has been used. Furthermore, inserting Eq. (23) into the second and third formulae of Eq. (15) and after some lengthy manipulations, we obtain

$$\left\{T_n + \frac{h}{r_0}\left[\left(\frac{\rho^s\Omega^2}{\rho} + \frac{2c_{66}^s}{c_{44}} - \frac{k_1^s}{c_{44}}n(n+1)\right)V_n - \frac{k_1^s + k_2^s}{c_{44}}W_n + \frac{e_{33}\gamma^s}{c_{44}\alpha^s}Z_n\right]\right\}\frac{\partial S_n^m(\theta,\varphi)}{\partial \theta}$$

$$-\left\{Y_n - \frac{h}{r_0}\left[\frac{\rho^s\Omega^2}{\rho} + \frac{c_{66}^s}{c_{44}}(2 - n(n+1))\right]U_n\right\}\frac{\partial S_n^m(\theta,\varphi)}{\sin\theta\partial\varphi} = 0, \quad (\xi = 1,\ n \geq 1)$$

$$\left\{T_n + \frac{h}{r_0}\left[\left(\frac{\rho^s\Omega^2}{\rho} + \frac{2c_{66}^s}{c_{44}} - \frac{k_1^s}{c_{44}}n(n+1)\right)V_n - \frac{k_1^s + k_2^s}{c_{44}}W_n + \frac{e_{33}\gamma^s}{c_{44}\alpha^s}Z_n\right]\right\}\frac{\partial S_n^m(\theta,\varphi)}{\sin\theta\partial\varphi}$$

$$+\left\{Y_n - \frac{h}{r_0}\left[\frac{\rho^s\Omega^2}{\rho} + \frac{c_{66}^s}{c_{44}}(2 - n(n+1))\right]U_n\right\}\frac{\partial S_n^m(\theta,\varphi)}{\partial\theta} = 0, \quad (\xi = 1,\ n \geq 1)$$
(26)

where the identity $\nabla_1^2 S_n^m(\theta,\varphi) + n(n+1)S_n^m(\theta,\varphi) = 0$ along with the relation $k_1^s - k_2^s = c_{11}^s - c_{12}^s = 2c_{66}^s$ has also been utilized. By invoking the following relation [47]

$$\sin\theta\frac{dP_n^m(\cos\theta)}{d\theta} = \frac{1}{2n+1}\left[n(n-m+1)P_{n+1}^m(\cos\theta) - (n+1)(n+m)P_{n-1}^m(\cos\theta)\right], \quad (27)$$

and the orthogonality of the associated Legendre functions with respect to the weight $\sin\theta$ [47], we can obtain from Eq. (26)

$$Y_n - \frac{h}{r_0}\left\{\frac{\rho^s\Omega^2}{\rho} + \frac{c_{66}^s}{c_{44}}[2 - n(n+1)]\right\}U_n = 0, \quad (\xi = 1,\ n \geq 1)$$

$$T_n + \frac{h}{r_0}\left\{\left[\frac{\rho^s\Omega^2}{\rho} + \frac{2c_{66}^s}{c_{44}} - \frac{k_1^s}{c_{44}}n(n+1)\right]V_n - \frac{k_1^s + k_2^s}{c_{44}}W_n + \frac{e_{33}\gamma^s}{c_{44}\alpha^s}Z_n\right\} = 0, \quad (\xi = 1,\ n \geq 1)$$
(28)

Equations (25) and (28) are the effective boundary conditions in terms of functions $U_n$, $V_n$, $W_n$ and $X_n$ for the piezoelectric nanosphere, which can be separated into two categories: one is related to the function $U_n$ in the first one of Eq. (28), and the other is expressed by the other three functions $V_n$, $W_n$ and $X_n$ in Eq. (25) and the second one of Eq. (28). If the surface effect is neglected, the effective boundary conditions (25) and (28) reduce to the classical boundary conditions [47].

Using the results obtained above, the free vibration of a piezoelectric nanosphere with surface effect may be divided into two independent classes, just like the macroscopic case without surface effect [47]. The first class $(n \geq 1)$, governed by the first one of Eq. (19) and the first one of Eq. (28), corresponds to an equi-volumetric motion of the piezoelectric nanosphere without the radial displacement and the electric potential, while for the second class $(n \geq 0)$ defined by the last three formulae of Eq. (19), Eq. (25) and the second one of Eq. (28), the mechanical displacements possess in general both transverse and radial components except for the case of $n = 0$, in which there exists only the radial component, but the rotation vector has no radial component [47].

### 5.3. *Frequency equation of the first class* $(n \geq 1)$

Substituting Eq. (20) into the first one of Eq. (28), we obtain

$$\left(\eta - \frac{3}{2}\right)J_\eta(\Omega) - \Omega J_{\eta+1}(\Omega) - \frac{h}{r_0}\left\{\frac{\rho^s \Omega^2}{\rho} + \frac{c_{66}^s}{c_{44}}\left[2 - n(n+1)\right]\right\}J_\eta(\Omega) = 0. \qquad (29)$$

This gives the frequency equation of the first class of vibration of the piezoelectric nanosphere with surface effect. In particular, when $n = 1$, the dispersion equation (29) corresponds to a torsional or rotary mode of the piezoelectric nanosphere. It can be seen that the frequency equation (29) contains no parameter related to the electric field, and hence the piezoelectric effect has completely no influence on the natural frequencies of the first class of vibration just as the macroscopic case without surface effect [47]. However, the natural frequencies are indeed influenced by the surface mass density and the surface elastic constants.

**5.4. *Frequency equation of the second class ($n \geq 0$)***

For $n = 0$, it has been verified [53,54] that one of the two solutions in Eq. (22) (e.g., the solution corresponding to $C_{02}$) will give no contribution to the stresses and electric displacements so that one has $C_{02} = 0$ in Eq. (22). Additionally, it also has been demonstrated [53,54] that the radial electric displacement component $D_r$ corresponding to the solution with the unknown $C_{01}$ will be zero. Therefore, the second one of the effective boundary conditions (25) is satisfied automatically. Substitution of Eq. (22) into the first one of Eq. (25) leads to the following frequency equation:

$$\frac{2W_{01}}{\xi} + \frac{f_4}{f_3}W'_{01} + \frac{f_8}{f_3}X'_{01} + \frac{h}{r_0}\left[\frac{2(k_1^s + k_2^s)}{c_{13}} - \frac{\rho^s \Omega^2 c_{44}}{\rho c_{13}}\right]W_{01} = 0, \quad (\xi = 1) \quad (30)$$

As described above, the frequency equation (30) corresponds to the purely radial vibration mode of the piezoelectric nanosphere with surface effect.

For $n \geq 1$, substituting Eq. (21) into Eq. (25) and the second one of Eq. (28), we can obtain a set of linear homogeneous algebraic equations for the undetermined constants $C_{nj}$. As a necessary condition for nontrivial solutions, the determinant of the coefficient matrix of the algebraic equations should vanish. This gives the frequency equation of the second class of vibration as follows

$$|N_{ij}|=0, \quad (i,j=1-3, \ \xi=1) \tag{31}$$

with

$$N_{1j} = R_{nj} + \frac{h}{r_0}\left\{\frac{k_1^s+k_2^s}{c_{13}}\left[2W_{nj}+n(n+1)V_{nj}\right]-\frac{\rho^s\Omega^2 c_{44}}{\rho c_{13}}W_{nj}-\frac{2e_{33}\gamma^s}{c_{13}\alpha^s}Z_{nj}\right\},$$

$$N_{2j} = \left(1-\frac{2h}{r_0}\right)Z_{nj}-\frac{h}{r_0}\frac{k_3^s}{\varepsilon_{33}}n(n+1)X_{nj}, \quad (j=1-3) \tag{32}$$

$$N_{3j} = T_{nj} + \frac{h}{r_0}\left\{\left[\frac{\rho^s\Omega^2}{\rho}+\frac{2c_{66}^s}{c_{44}}-\frac{k_1^s}{c_{44}}n(n+1)\right]V_{nj}-\frac{k_1^s+k_2^s}{c_{44}}W_{nj}+\frac{e_{33}\gamma^s}{c_{44}\alpha^s}Z_{nj}\right\},$$

where

$$\begin{aligned}
R_{nj}(\xi) &= 2W_{nj}/\xi + V_{nj}n(n+1)/\xi + (f_4/f_3)W'_{nj} + (f_8/f_3)X'_{nj}, \\
T_{nj}(\xi) &= W_{nj}/\xi - V'_{nj} + V_{nj}/\xi + f_5 f_8 X_{nj}/\xi, \\
Z_{nj}(\xi) &= 2f_6 W_{nj}/\xi + f_6 V_{nj}n(n+1)/\xi + W'_{nj} - X'_{nj}
\end{aligned} \tag{33}$$

Note that if the surface effect is excluded, the frequency equations (29)-(31) will reduce to the classical ones for a macroscopic piezoelectric sphere [47]. Again, it should be emphasized here that the frequency equations (29)-(31) are derived for the case that the TS effect is discarded. When the TS effect is considered, the frequency equations can be derived in a similar manner but it is omitted here for simplicity. The TS effect on the free vibration of piezoelectric nanospheres will be investigated numerically in the following section.

## 6. Numerical results and discussions

In this section, in order to quantitatively investigate the surface effects on the free vibration behavior of piezoelectric nanospheres with different sizes, numerical calculations are conducted for a PZT-5H nanosphere with a radial polarization. The used bulk (or macroscopic) material constants for PZT-5H [47] are given by: $c_{11}=126$ GPa, $c_{12}=79.5$ GPa, $c_{13}=84.1$ GPa, $c_{33}=117$ GPa, $c_{44}=23$ GPa,

$e_{15} = 17$ C/m$^2$, $e_{31} = -6.5$ C/m$^2$, $e_{33} = 23.3$ C/m$^2$, $\kappa_{11} = 1.5 \times 10^{-8}$ C/Vm, $\kappa_{33} = 1.3 \times 10^{-8}$ C/Vm and $\rho = 7500$ kg/m$^3$.

Although the fabrication and synthesis of the PZT nanoparticles/nanospheres have been conducted utilizing different approaches, as mentioned previously, the experimental characterization of their surface properties is still lacking. On the other hand, although atomistic simulations including molecular dynamics simulations (MDS) and quantum mechanics calculations have been employed to determine the surface piezoelectric properties of ZnO, BaTiO3, GaN, and AlN [17,56,57], no validated data for the PZT-5H nanostructures can be found in the literature, possibly because of its more complex atomic structure. It is also worth mentioning that the definition formula (16) for the differentiation of the effective electric field with respect to the electric field in [57] is questionable unless the electric field itself is uniform in the nanostructure. Therefore, we take the estimated values of the surface material constants in the HY theory as [28-32]: $\bar{c}_{11}^Y = 7.56$ N/m, $\bar{e}_{31}^Y = -3 \times 10^{-8}$ C/m, $\bar{\kappa}_{11}^Y = 1.505 \times 10^{-17}$ C/V and $\bar{\rho}_0^Y = 7.5 \times 10^{-6}$ kg/m$^2$. These material constants have been adopted previously to investigate the surface piezoelectric effect on the static and dynamic electromechanical properties of PZT-5H nanostructures [29-32]. Other surface material parameters can be assumed as $\bar{c}_{ij}^Y = (\bar{c}_{11}^Y / c_{11}) c_{ij}$, $\bar{e}_{ij}^Y = (\bar{e}_{31}^Y / e_{31}) e_{ij}$ and $\bar{\kappa}_{ij}^Y = (\bar{\kappa}_{11}^Y / \kappa_{11}) \kappa_{ij}$ following the treatment of Yan and Jiang [30]. Furthermore, since the lattice constant of PZT is 3.99 Å (~0.4 nm) [31,58] and the surface region typically involves a few atomic layers, the thickness of the surface spherical shell can be approximately taken as 1 nm in our calculation. Therefore, the surface material parameters (with superscript *s*) in our surface piezoelectricity theory can be evaluated from Eq. (16) according to the HY surface material parameters. It is worth noting that the surface material parameters $\beta^s/\alpha^s$ and $e_{15}^s/c_{44}^s$ appearing in the numerical calculations concerning the TS effect can be determined from the second and fifth formulae of Eq. (16).

For the sake of brevity, in the following discussions, we introduce three abbreviations below, namely Classical, HY and PTS, to denote the classical piezoelectricity theory without surface effect, the HY surface piezoelectricity theory without TS effect, and the proposed surface piezoelectricity theory with TS effect, respectively.

It should be mentioned that the frequency equations (29)-(31) are derived for 3D motion and hence can predict an infinite number of eigen-frequencies. However, in what follows, we only consider the smallest positive natural or eigen-frequency that is of practical significance [47,53,54].

**6.1. *Surface effect on the vibration of the first class*** $(n \geq 1)$

In this subsection, we will numerically investigate the surface effect on the free vibration of the first class of the PZT-5H nanosphere using three different theories (i.e., Classical, HY, and PTS).

The curves of the lowest non-dimensional natural frequency $\Omega$ of the first class versus the mode number $n$ are depicted in Figure 1 for four different values of the radius $r_0$ of the piezoelectric nanosphere, wherein Figure 1a displays the HY results while Figure 1b shows the PTS results. The Classical results are also included in the two figures for comparison. One can observe here that, for the piezoelectric nanosphere with a given radius $r_0$, the lowest natural frequencies for the torsional mode $(n=1)$ are higher than those of the first-order non-torsional mode $(n=2)$, and the frequency of the non-torsional mode increases with the increase of the mode number. It can be seen from Figure 1 that the natural frequencies of the first class are reduced when the surface effect is considered. With the decrease of the radius $r_0$ of the bulk core, the natural frequencies deviate much more from those without surface effect. In addition, we observe from Figure 1 that the surface effect on the natural frequencies of the first class is more significant for high-order modes than those for low-order modes. In fact, the frequencies of the high-order modes are considerably lowered. Comparing Figure 1b with Figure 1a, we can also find that the predictions from the PTS theory agree quite well with those from the HY theory for a relatively large radius, while for the piezoelectric nanosphere with a smaller radius, the PTS theory with the TS effects included will generally make the deviation from the Classical results more obvious than the HY theory.

The variations of the natural frequency differences $\Delta\Omega_1 = \Omega_h - \Omega_c$ and $\Delta\Omega_2 = \Omega_t - \Omega_h$ versus the mode number are shown in Figure 2 for two different radii $r_0 = 10$ nm and $5$ nm. Here, $\Omega_c$ is the frequency without surface effect, and $\Omega_h$ and $\Omega_t$ are the frequencies predicted from the HY and the PTS theory, respectively. As shown in Figure 2, the absolute values of the frequency difference for low-order modes are smaller than those for high-order modes, which means that the surface effect has a more significant influence on the frequency of high-order modes. Besides, the absolute values of the frequency differences $\Delta\Omega_1$ and $\Delta\Omega_2$ increase with the decrease of the radius $r_0$. These observations are consistent with those in Figure 1. It is also noted from Figure 2 that the PTS results are different from those predicted by the HY theory, especially for high-order modes.

The curves of the lowest natural frequency for $n=1$ and $n=2$ versus $\log_{10}[r_0(\text{nm})]$ are displayed in Figure 3 for both HY and PTS theories in order to clearly show the surface effect on the free vibration of the first class. It is seen that the frequencies predicted by the HY and PTS theories for both $n=1$ and $n=2$ approach those without surface effect (i.e., $\Omega=5.7635$ and 2.5141, respectively) when the bulk core radius $r_0$ of the nanosphere becomes infinitely large, which is expected since the surface effect becomes negligible for a macroscopic piezoelectric sphere. However, it can be found from Figure 3 that the frequencies for both $n=1$ and $n=2$ begin to deviate from those without surface effect at $\log_{10} r_{0c}=2.24$, i.e., $r_{0c}=174$ nm, which may be regarded as a critical radius. Thus, when the radius decreases to the critical radius $r_{0c}$, the surface effect may be significant and should be taken into account in the modeling [19,31,32,46]. Furthermore, we note that the frequencies predicted by the PTS theory, which takes account of the TS effect, begin to deviate from those by the HY theory at $\log_{10} r_{0t}=1.28$ (i.e., $r_{0t}=19$ nm), below which the TS effect may play an important role in the free vibration of the piezoelectric nanosphere. This phenomenon is similar to that reported by Chen et al. [19]. Another interesting phenomenon is that the frequency for $n=2$ predicted by the PTS theory reduces to zero, when $\log_{10} r_0$ decreases approximately to the value 0.4413 (i.e., $r_0 = 3$ nm), below which no real frequencies could be found along the continuation of the corresponding frequency branch (i.e., the lowest branch).

**6.2. *Surface effect on the vibration of the second class*** $(n \geq 0)$

Now we turn to the free vibration of the second class. As mentioned earlier, the piezoelectric nanosphere vibrates only in the radial direction when $n = 0$, which is usually referred to as the breathing mode. The variations of the lowest natural frequency for the breathing mode with $\log_{10}[r_0(\text{nm})]$ are shown in Figure 4a for both HY and PTS theories. Just like in Figure 3, the surface effect on the breathing mode is evident when the radius decreases to $\log_{10} r_{0c} = 2.16$ (i.e., the critical radius $r_{0c} = 145$ nm), and both the HY and PTS predictions approach the classical one (i.e., $\Omega = 6.8407$) when the radius gets infinitely large. There also exists a value $\log_{10} r_{0t} = 1.2$ (i.e., $r_{0t} = 16$ nm), below which the TS effect on the breathing mode of the piezoelectric nanosphere should be considered, which is similar to the phenomenon shown in Figure 3.

Furthermore, the variations of the natural frequency differences $\Delta\Omega_1 = \Omega_h - \Omega_c$ and $\Delta\Omega_2 = \Omega_t - \Omega_h$ for the breathing mode versus $\log_{10}[r_0(\text{nm})]$ are plotted in Figure 4b, where $\Omega_c$, $\Omega_h$ and $\Omega_t$ have the same meaning as introduced before. It should be noted here that the surface effect and the TS effect on the breathing mode of the piezoelectric nanosphere will become remarkable when the radius reduces to the same values $r_{0c} = 145$ nm and $r_{0t} = 16$ nm, respectively, as those in Figure 4a. It is also worth mentioning that, after $\log_{10} r_0$ decreases approximately to the value 0.22 (i.e., $r_0 = 1.7$ nm), the absolute value of the frequency difference $\Delta\Omega_2$ is larger than that of $\Delta\Omega_1$, which means that the TS effect is crucial to the frequency of the breathing mode.

Now let's turn our attention to the non-breathing modes $(n \geq 1)$ of the second class of the vibration of the piezoelectric nanosphere. The curves of the lowest natural frequencies for the non-breathing modes $n = 1$ and $n = 2$ versus $\log_{10}[r_0(\text{nm})]$ are depicted in Figure 5 for both HY and PTS theories. Similar to Figures 3 and 4a, the natural frequencies predicted by HY and PTS theories for both $n = 1$ and $n = 2$ approach the classical values (i.e., $\Omega = 4.2951$ and 2.8818, respectively) when the radius $r_0$ tends to infinity. The surface effect on the non-breathing modes $n = 1$ and $n = 2$ begins to play an important role when the radius of the nanosphere decreases to $\log_{10} r_{0c} = 1.88$ and 1.96 (corresponding to $r_{0c} = 76$ nm and 91 nm, respectively). In addition, the TS effect on the non-breathing modes $n = 1$ and $n = 2$ of the piezoelectric nanosphere should be taken into account when the radius arrives at the values $\log_{10} r_{0t} = 1.16$ and 1.24 (i.e., $r_{0t} = 14.5$ nm and 17 nm, respectively). It should be pointed out that, for the non-breathing modes $n = 1$ and $n = 2$, the frequencies predicted by the PTS theory also decrease to zero at $\log_{10} r_0 = 0.1$ and 0.389 (i.e., $r_0 = 1.3$ nm and 2.45 nm, respectively).

The variations of the natural frequency differences $\Delta\Omega_1$ and $\Delta\Omega_2$ for the non-breathing modes $n = 1$ and $n = 2$ versus $\log_{10}[r_0(\text{nm})]$ are plotted in Figure 6. We can obtain similar results to those in Figure 4b. Particularly, the absolute value of the frequency difference $\Delta\Omega_2$ for $n = 2$ is larger than that of $\Delta\Omega_1$ in Figure 6b after $\log_{10} r_0$ decreases approximately to the value 0.5 (i.e., $r_0 = 3.2$ nm). Below this particular sphere radius, the TS effect on the non-breathing mode $n = 2$ is significant to the frequency of the piezoelectric nanosphere.

## 7. Conclusions

In this study, the surface effect on the free vibration of spherically isotropic piezoelectric nanospheres with a radial polarization is analyzed. A theory of surface piezoelectricity (i.e., the effective boundary conditions) for a spherical material boundary is presented by means of the state-space formalism. The proposed first-order surface piezoelectricity theory contains three different contributions, including the electroelastic effect, the inertia effect, and the transverse stress effect. In the absence of residual stresses and electric displacements, the established theory will give rise to the same governing equations as the HY surface piezoelectricity theory, provided that the TS effect in the first-order approximation is omitted and the surface material properties are properly defined. Based on the derived effective boundary conditions, the exact frequency equations for the two distinct classes of free vibration are obtained, which can well predict the size-dependent frequencies of piezoelectric nanospheres with surface effect.

Numerical results for a PZT-5H nanosphere are presented to illustrate that the surface effect, especially the TS effect which is neglected in the HY surface piezoelectricity theory, has a particularly significant influence on the natural frequencies of the PZT-5H nanosphere. Specifically, the surface effect on the natural frequencies of the first class is more significant for high-order modes than those for low-order modes. The natural frequencies of the vibration of the first class for both $n=1$ and $n=2$ become obviously size-dependent when the radius of the PZT-5H nanosphere decreases to the critical radius $r_{0c}=174\,\text{nm}$. For the breathing mode $n=0$ and non-breathing modes $n=1$ and $n=2$ of the vibration of the second class, the critical radii, below which the surface effect is evident, are $r_{0c}=145\,\text{nm}$, $76\,\text{nm}$ and $91\,\text{nm}$, respectively. In addition, the TS effect may play a significant role in the free vibration of the PZT-5H nanosphere below the radius $r_{0t}=19\,\text{nm}$ for the modes $n=1$ and $n=2$ of the vibration of the first class. For the breathing mode $n=0$ and non-breathing modes $n=1$ and $n=2$ of the vibration of the second class, the radii, below which the TS effect is important and should be taken into account in the modeling, are $r_{0t}=16\,\text{nm}$, $14.5\,\text{nm}$ and $17\,\text{nm}$, respectively. The aforementioned observations are important for a better understanding of the dynamic behavior of the piezoelectric nanospheres.

This work provides not only a novel method to derive the governing equations for surface piezoelectricity theory for a spherical material boundary, but also the analytical frequency equations of piezoelectric nanospheres. Based on the present formulations, more accurate predictions on the size-dependent vibration frequencies for piezoelectric nanospherical devices in NEMS can be obtained.

**Acknowledgments**

The work was supported by the National Natural Science Foundation of China (Nos. 11621062 and 11532001), the German Research Foundation (DFG, Project-No: ZH 15/20-1), and the China Scholarship Council (CSC). Partial support from the Fundamental Research Funds for the Central Universities (No. 2016XZZX001-05) is also acknowledged.

**Appendix A: Elements of the system matrix M**

The nonzero elements of the system matrix $\mathbf{M}$ in the state equation (4) are given as follows:

$$M_{1111} = -\frac{2\beta}{\alpha}\frac{1}{r},\ M_{1112} = -\frac{\beta}{\alpha}\frac{1}{r}\left(\frac{\partial}{\partial\theta}+\cot\theta\right),\ M_{1113}=M_{2231}=-\frac{\beta}{\alpha}\frac{1}{r\sin\theta}\frac{\partial}{\partial\varphi},\ M_{1121}=-\frac{\partial}{r\partial\theta},$$

$$M_{1122}=M_{1133}=\frac{1}{r},\ M_{1124}=-\frac{e_{15}}{c_{44}}\frac{\partial}{r\partial\theta},\ M_{1131}=M_{2213}=-\frac{1}{r\sin\theta}\frac{\partial}{\partial\varphi},\ M_{1134}=M_{2243}=-\frac{e_{15}}{c_{44}}\frac{1}{r\sin\theta}\frac{\partial}{\partial\varphi},$$

$$M_{1141}=-M_{2214}=\frac{2\gamma}{\alpha}\frac{1}{r},\ M_{1142}=\frac{\gamma}{\alpha}\frac{1}{r}\left(\frac{\partial}{\partial\theta}+\cot\theta\right),\ M_{1143}=M_{2234}=\frac{\gamma}{\alpha}\frac{1}{r\sin\theta}\frac{\partial}{\partial\varphi},$$

$$M_{1211}=\frac{\kappa_{33}}{\alpha},\ M_{1214}=M_{1241}=\frac{e_{33}}{\alpha},\ M_{1222}=M_{1233}=\frac{1}{c_{44}},\ M_{1244}=-\frac{c_{33}}{\alpha},$$

$$M_{2111}=\rho\frac{\partial^2}{\partial t^2}+\frac{2(k_1+k_2)}{r^2},\ M_{2112}=\frac{k_1+k_2}{r^2}\left(\cot\theta+\frac{\partial}{\partial\theta}\right),\ M_{2121}=-\frac{k_1+k_2}{r^2}\frac{\partial}{\partial\theta},$$

$$M_{2113}=-M_{2131}=\frac{k_1+k_2}{r^2}\frac{1}{\sin\theta}\frac{\partial}{\partial\varphi},\ M_{2122}=\rho\frac{\partial^2}{\partial t^2}-\frac{1}{r^2}\left(k_1\nabla_4^2-k_2+\frac{c_{66}}{\sin^2\theta}\frac{\partial^2}{\partial\varphi^2}\right),$$

$$M_{2123}=-\frac{1}{r^2\sin\theta}\left[(k_2+c_{66})\frac{\partial^2}{\partial\varphi\partial\theta}-(k_1+c_{66})\cot\theta\frac{\partial}{\partial\varphi}\right],$$

$$M_{2132}=-\frac{1}{r^2\sin\theta}\left[(c_{66}+k_2)\frac{\partial^2}{\partial\theta\partial\varphi}+(c_{66}+k_1)\cot\theta\frac{\partial}{\partial\varphi}\right],$$

$$M_{2133}=\rho\frac{\partial^2}{\partial t^2}-\frac{1}{r^2}\left[\frac{k_1}{\sin^2\theta}\frac{\partial^2}{\partial\varphi^2}+c_{66}\left(\nabla_4^2+1\right)\right],\ M_{2144}=\frac{k_3}{r^2}\nabla_1^2,$$

$$M_{2211}=2(\frac{\beta}{\alpha}-1)\frac{1}{r},\ M_{2212}=-\frac{1}{r}(\frac{\partial}{\partial\theta}+\cot\theta),\ M_{2221}=-\frac{\beta}{\alpha}\frac{\partial}{r\partial\theta},\ M_{2222}=M_{2233}=-\frac{3}{r},$$

$$M_{2224}=\frac{\gamma}{\alpha}\frac{\partial}{r\partial\theta},\ M_{2242}=-\frac{e_{15}}{c_{44}}\left(\frac{\partial}{\partial\theta}+\cot\theta\right)\frac{1}{r},\ M_{2244}=-\frac{2}{r}$$

where $M_{ijkl}$ represents the element on the *k*th row and *l*th column of the sub-matrices $\mathbf{M}_{ij}$ and

$$\alpha=c_{33}\kappa_{33}+e_{33}^2,\ \beta=c_{13}\kappa_{33}+e_{33}e_{31},\ \gamma=c_{33}e_{31}-e_{33}c_{13},$$

$$k_1=c_{11}-\frac{c_{13}\beta}{\alpha}+\frac{e_{31}\gamma}{\alpha},\ k_2=c_{12}-\frac{c_{13}\beta}{\alpha}+\frac{e_{31}\gamma}{\alpha},\ k_3=\frac{e_{15}^2}{c_{44}}+\kappa_{11},$$

$$\nabla_1^2=\frac{\partial^2}{\partial\theta^2}+\cot\theta\frac{\partial}{\partial\theta}+\frac{1}{\sin^2\theta}\frac{\partial^2}{\partial\varphi^2},\ \nabla_4^2=\frac{\partial^2}{\partial\theta^2}+\cot\theta\frac{\partial}{\partial\theta}-\cot^2\theta$$

**Appendix B: Equations for the HY surface piezoelectricity in spherical**

**coordinates**

Huang and Yu [28] generalized the idea of the GM theory to establish a surface piezoelectricity model for piezoelectric nanomaterials. In the absence of residual stresses and electric displacements, the generalized geometric, constitutive and equilibrium equations in the HY theory [28] in tensor notation are given by

$$\mathbf{E}^Y = -\mathrm{grad}\Phi, \quad \boldsymbol{\varepsilon}^Y = \frac{1}{2}\left(\mathbf{D}_S\mathbf{u} + \mathbf{D}_S^T\mathbf{u}\right), \quad \mathbf{D}_S = \mathbf{P}(\mathrm{grad}_S\mathbf{u}),$$
$$\boldsymbol{\Sigma} = \overline{\mathbf{c}}^Y : \boldsymbol{\varepsilon}^Y - \left(\overline{\mathbf{e}}^Y\right)^T \mathbf{E}^Y, \quad \mathbf{D}^Y = \overline{\mathbf{e}}^Y : \boldsymbol{\varepsilon}^Y + \overline{\boldsymbol{\kappa}}^Y \mathbf{E}^Y, \tag{B1}$$
$$\mathrm{div}_S\boldsymbol{\Sigma} = \boldsymbol{\sigma}\mathbf{n} + \overline{\rho}_0^Y \ddot{\mathbf{u}}, \quad \mathrm{div}_S\mathbf{D}^Y = \mathbf{Dn}$$

where $\overline{\rho}_0^Y$ is the surface mass density; $\overline{\mathbf{c}}^Y$, $\overline{\mathbf{e}}^Y$ and $\overline{\boldsymbol{\kappa}}^Y$ are the surface elastic, piezoelectric, and dielectric tensors, respectively, in the HY theory; $\boldsymbol{\Sigma}$ and $\boldsymbol{\varepsilon}^Y$ denote the surface stress and surface strain tensors; $\mathbf{D}^Y$ and $\mathbf{E}^Y$ denote the surface electric displacement and surface electric field vectors; $\mathbf{P}$ and $\mathbf{n}$ are the projection tensor onto the tangential space of the surface and the outward unit normal to the surface; and the operators $\mathrm{grad}$, $\mathrm{grad}_S$ and $\mathrm{div}_S$ are defined in [21,59].

The displacement vector in spherical coordinates $(r,\theta,\varphi)$ can be written as $\mathbf{u} = u_r\mathbf{e}_r + u_\theta\mathbf{e}_\theta + u_\varphi\mathbf{e}_\varphi$ with $\mathbf{e}_i$ being the unit vector along the $i$-axis ($i = r,\theta,\varphi$). Thus, the projection tensor onto the tangential space is $\mathbf{P} = \mathbf{I} - \mathbf{e}_r\mathbf{e}_r$, where $\mathbf{I}$ is the 3D identity tensor. In accordance with the definitions in [59], by setting $v_1 = \theta$, $v_2 = \varphi$ and $v_3 = r$, the corresponding metric coefficients are found as $h_1 = r$, $h_2 = r\sin\theta$ and $h_3 = 1$.

Accordingly, the first and second formulae of Eq. (B1) yield the following non-zero components of the surface strain tensor and the tangential components of the surface electric field vector

$$\varepsilon_{11}^Y = \frac{1}{r}\left(\frac{\partial u_\theta}{\partial \theta}+u_r\right), \quad \varepsilon_{12}^Y = \varepsilon_{21}^Y = \frac{1}{2r}\left(\frac{\partial u_\varphi}{\partial \theta}+\frac{1}{\sin\theta}\frac{\partial u_\theta}{\partial \varphi}-u_\varphi\cot\theta\right),$$
$$\varepsilon_{22}^Y = \frac{1}{r\sin\theta}\left(\frac{\partial u_\varphi}{\partial \varphi}+u_\theta\cos\theta+u_r\sin\theta\right), \quad E_1^Y = -\frac{1}{r}\frac{\partial \Phi}{\partial \theta}, \quad E_2^Y = -\frac{1}{r\sin\theta}\frac{\partial \Phi}{\partial \varphi}$$
(B2)

For spherically isotropic piezoelectric nanospheres with a radial polarization, substitution of Eq. (B2) into the fourth and fifth formulae of Eq. (B1) leads to the following surface stress and surface electric displacement components

$$\Sigma_{11} = \frac{\bar{c}_{11}^Y}{r}\left(\frac{\partial u_\theta}{\partial \theta}+u_r\right)+\frac{\bar{c}_{12}^Y}{r\sin\theta}\left(\frac{\partial u_\varphi}{\partial \varphi}+u_\theta\cos\theta+u_r\sin\theta\right)-\bar{e}_{31}^Y E_3^Y,$$
$$\Sigma_{22} = \frac{\bar{c}_{12}^Y}{r}\left(\frac{\partial u_\theta}{\partial \theta}+u_r\right)+\frac{\bar{c}_{11}^Y}{r\sin\theta}\left(\frac{\partial u_\varphi}{\partial \varphi}+u_\theta\cos\theta+u_r\sin\theta\right)-\bar{e}_{31}^Y E_3^Y,$$
$$\Sigma_{12} = \Sigma_{21} = \frac{\bar{c}_{66}^Y}{r}\left(\frac{\partial u_\varphi}{\partial \theta}+\frac{1}{\sin\theta}\frac{\partial u_\theta}{\partial \varphi}-u_\varphi\cot\theta\right), \quad (B3)$$
$$D_1^Y = -\frac{\bar{\kappa}_{11}^Y}{r}\frac{\partial \Phi}{\partial \theta}, \quad D_2^Y = -\frac{\bar{\kappa}_{11}^Y}{r\sin\theta}\frac{\partial \Phi}{\partial \varphi},$$
$$D_3^Y = \frac{\bar{e}_{31}^Y}{r}\left(\frac{\partial u_\theta}{\partial \theta}+u_r\right)+\frac{\bar{e}_{31}^Y}{r\sin\theta}\left(\frac{\partial u_\varphi}{\partial \varphi}+u_\theta\cos\theta+u_r\sin\theta\right)+\bar{\kappa}_{33}^Y E_3^Y$$

Thus, from the last one of Eq. (B3), we can obtain

$$E_3^Y = -\frac{1}{\bar{\kappa}_{33}^Y}\left[\frac{\bar{e}_{31}^Y}{r}\left(\frac{\partial u_\theta}{\partial \theta}+u_r\right)+\frac{\bar{e}_{31}^Y}{r\sin\theta}\left(\frac{\partial u_\varphi}{\partial \varphi}+u_\theta\cos\theta+u_r\sin\theta\right)-D_3^Y\right] \quad (B4)$$

Upon substituting Eq. (B4) into the first and second formulae of Eq. (B3), the surface normal stresses become

$$\Sigma_{11} = \frac{\bar{k}_1^Y}{r}\left(\frac{\partial u_\theta}{\partial \theta}+u_r\right)+\frac{\bar{k}_2^Y}{r\sin\theta}\left(\frac{\partial u_\varphi}{\partial \varphi}+u_\theta\cos\theta+u_r\sin\theta\right)-\frac{\bar{e}_{31}^Y}{\bar{\kappa}_{33}^Y}D_3^Y,$$
$$\Sigma_{22} = \frac{\bar{k}_2^Y}{r}\left(\frac{\partial u_\theta}{\partial \theta}+u_r\right)+\frac{\bar{k}_1^Y}{r\sin\theta}\left(\frac{\partial u_\varphi}{\partial \varphi}+u_\theta\cos\theta+u_r\sin\theta\right)-\frac{\bar{e}_{31}^Y}{\bar{\kappa}_{33}^Y}D_3^Y$$
(B5)

where

$$\bar{k}_1^Y = \bar{c}_{11}^Y + \left(\bar{e}_{31}^Y\right)^2/\bar{\kappa}_{33}^Y, \quad \bar{k}_2^Y = \bar{c}_{12}^Y + \left(\bar{e}_{31}^Y\right)^2/\bar{\kappa}_{33}^Y \quad (B6)$$

Finally, making use of the definitions of the surface divergence $\text{div}_S$ for a vector and a tensor in the appendix of Ref. [59] and then substituting Eqs. (B3) and (B5) into the sixth and seventh formulae of Eq. (B1), we obtain for a piezoelectric spherical surface

$$\sigma_{rr} + \left[\bar{\rho}_0^Y \frac{\partial^2}{\partial t^2} + \frac{2(\bar{k}_1^Y + \bar{k}_2^Y)}{r_0^2}\right] u_r + \frac{\bar{k}_1^Y + \bar{k}_2^Y}{r_0^2}\left(\cot\theta + \frac{\partial}{\partial \theta}\right) u_\theta$$
$$+ \frac{\bar{k}_1^Y + \bar{k}_2^Y}{r_0^2 \sin\theta} \frac{\partial u_\varphi}{\partial \varphi} - \frac{2\bar{e}_{31}^Y}{\bar{\kappa}_{33}^Y r_0} D_r^Y = 0,$$

$$\sigma_{r\theta} - \frac{\bar{k}_1^Y + \bar{k}_2^Y}{r_0^2} \frac{\partial u_r}{\partial \theta} + \left[\bar{\rho}_0^Y \frac{\partial^2}{\partial t^2} - \frac{1}{r_0^2}\left(\bar{k}_1^Y \nabla_4^2 - \bar{k}_2^Y + \frac{\bar{c}_{66}^Y}{\sin^2\theta} \frac{\partial^2}{\partial \varphi^2}\right)\right] u_\theta$$
$$- \frac{1}{r_0^2 \sin\theta}\left[(\bar{k}_2^Y + \bar{c}_{66}^Y) \frac{\partial^2}{\partial \varphi \partial \theta} - (\bar{k}_1^Y + \bar{c}_{66}^Y)\cot\theta \frac{\partial}{\partial \varphi}\right] u_\varphi + \frac{\bar{e}_{31}^Y}{\bar{\kappa}_{33}^Y r_0} \frac{\partial D_r^Y}{\partial \theta} = 0,$$

$$\sigma_{r\varphi} - \frac{\bar{k}_1^Y + \bar{k}_2^Y}{r_0^2 \sin\theta} \frac{\partial u_r}{\partial \varphi} - \frac{1}{r_0^2 \sin\theta}\left[(\bar{k}_2^Y + \bar{c}_{66}^Y) \frac{\partial^2}{\partial \theta \partial \varphi} + (\bar{k}_1^Y + \bar{c}_{66}^Y)\cot\theta \frac{\partial}{\partial \varphi}\right] u_\theta$$
$$+ \left\{\bar{\rho}_0^Y \frac{\partial^2}{\partial t^2} - \frac{1}{r_0^2}\left[\frac{\bar{k}_1^Y}{\sin^2\theta} \frac{\partial^2}{\partial \varphi^2} + \bar{c}_{66}^Y(\nabla_4^2 + 1)\right]\right\} u_\varphi + \frac{\bar{e}_{31}^Y}{\bar{\kappa}_{33}^Y} \frac{1}{r_0 \sin\theta} \frac{\partial D_r^Y}{\partial \varphi} = 0,$$

$$D_r + \frac{\bar{\kappa}_{11}^Y}{r_0^2} \nabla_1^2 \Phi - \frac{2}{r_0} D_r^Y = 0$$

(B7)

where $r_0$ is the radius of the spherical core. Equation (B7) represents the governing equations of the HY theory for a spherically isotropic material surface of the piezoelectric nanospheres with a radial polarization. They can also be seen as the effective boundary conditions on the surface of a piezoelectric nanosphere.

# Figure Captions

**Figure 1.** Lowest non-dimensional natural frequencies $\Omega$ of the first class versus the mode number $n$ for four different radii $r_0$ of the piezoelectric nanosphere: (a) HY; (b) PTS.

**Figure 2.** Non-dimensional frequency differences $\Delta\Omega_1 = \Omega_h - \Omega_c$ and $\Delta\Omega_2 = \Omega_t - \Omega_h$ versus the mode number for two different radii $r_0 = 10$ nm and $5$ nm. Note: $\Omega_c$, $\Omega_h$ and $\Omega_t$ are the frequencies predicted from the Classical, HY and PTS theories, respectively.

**Figure 3.** Non-dimensional frequencies for modes $n=1$ and $n=2$ of the first class versus $\log_{10}[r_0(\text{nm})]$ for both HY and PTS theories.

**Figure 4.** (a) Non-dimensional frequencies for the breathing mode $n=0$ of the second class versus $\log_{10}[r_0(\text{nm})]$ for both HY and PTS theories; (b) Non-dimensional frequency differences $\Delta\Omega_1$ and $\Delta\Omega_2$ versus $\log_{10}[r_0(\text{nm})]$ for the breathing mode.

**Figure 5.** Non-dimensional frequencies for the non-breathing modes of the second class versus $\log_{10}[r_0(\text{nm})]$ for both HY and PTS theories: (a) $n=1$; (b) $n=2$.

**Figure 6.** Non-dimensional frequency differences $\Delta\Omega_1$ and $\Delta\Omega_2$ versus $\log_{10}[r_0(\text{nm})]$ for the non-breathing modes of the second class: (a) $n=1$; (b) $n=2$.

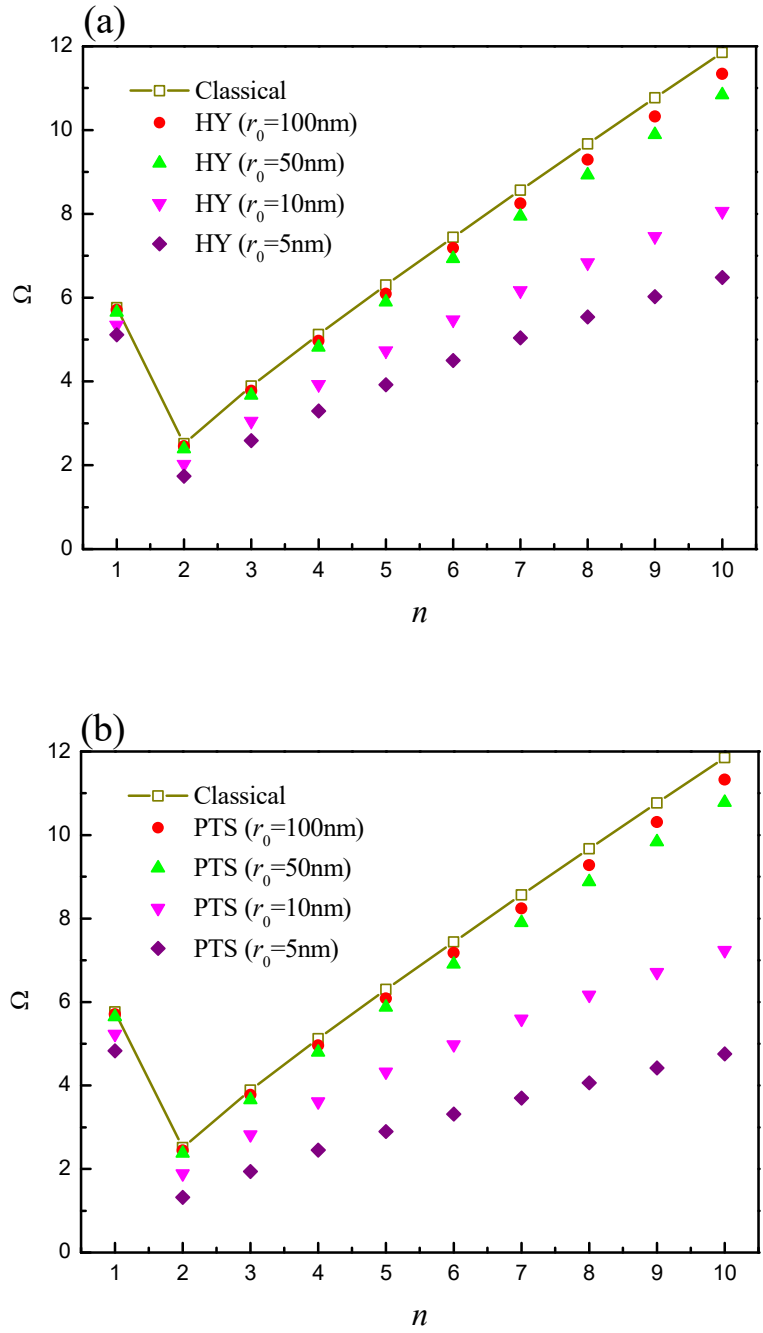

**Figure 1.** Lowest non-dimensional natural frequencies $\Omega$ of the first class versus the mode number $n$ for four different radii $r_0$ of the piezoelectric nanosphere: (a) HY; (b) PTS.

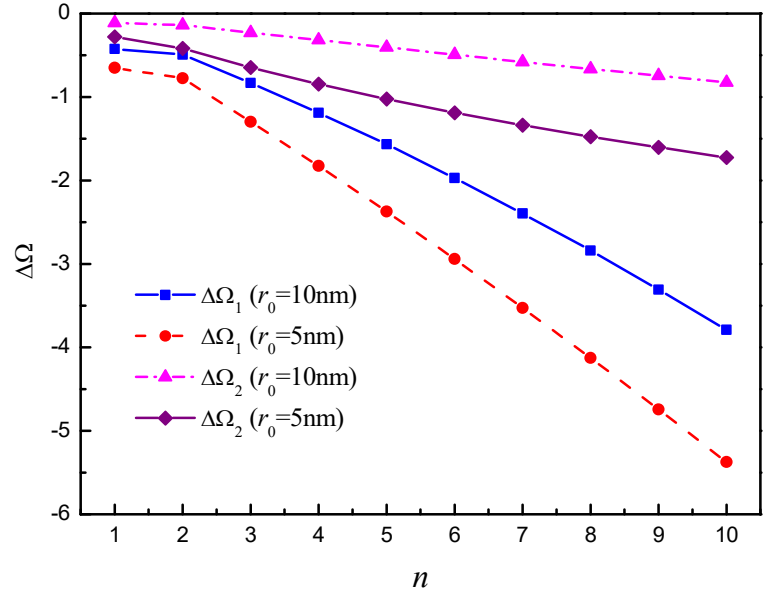

**Figure 2.** Non-dimensional frequency differences $\Delta\Omega_1 = \Omega_h - \Omega_c$ and $\Delta\Omega_2 = \Omega_t - \Omega_h$ versus the mode number for two different radii $r_0 = 10$ nm and $5$ nm. Note: $\Omega_c$, $\Omega_h$ and $\Omega_t$ are the frequencies predicted from the Classical, HY and PTS theories, respectively.

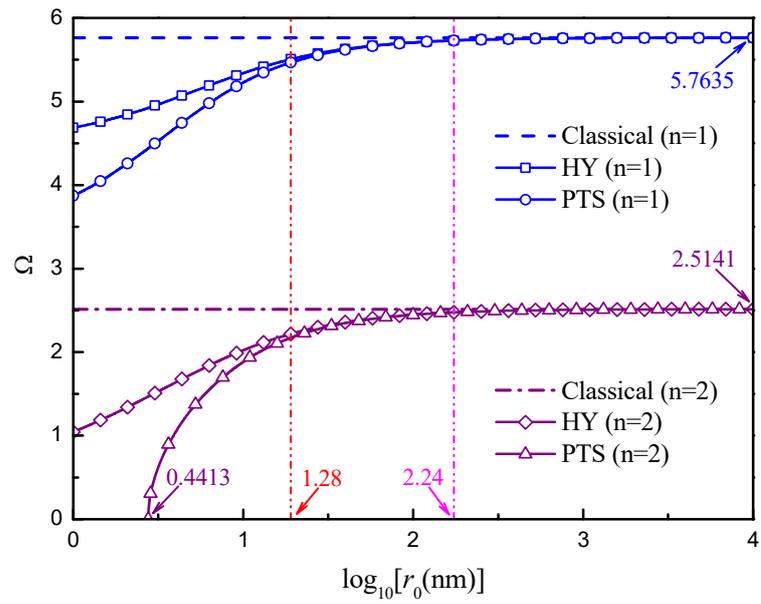

**Figure 3.** Non-dimensional frequencies for modes $n=1$ and $n=2$ of the first class versus $\log_{10}[r_0(\text{nm})]$ for both HY and PTS theories.

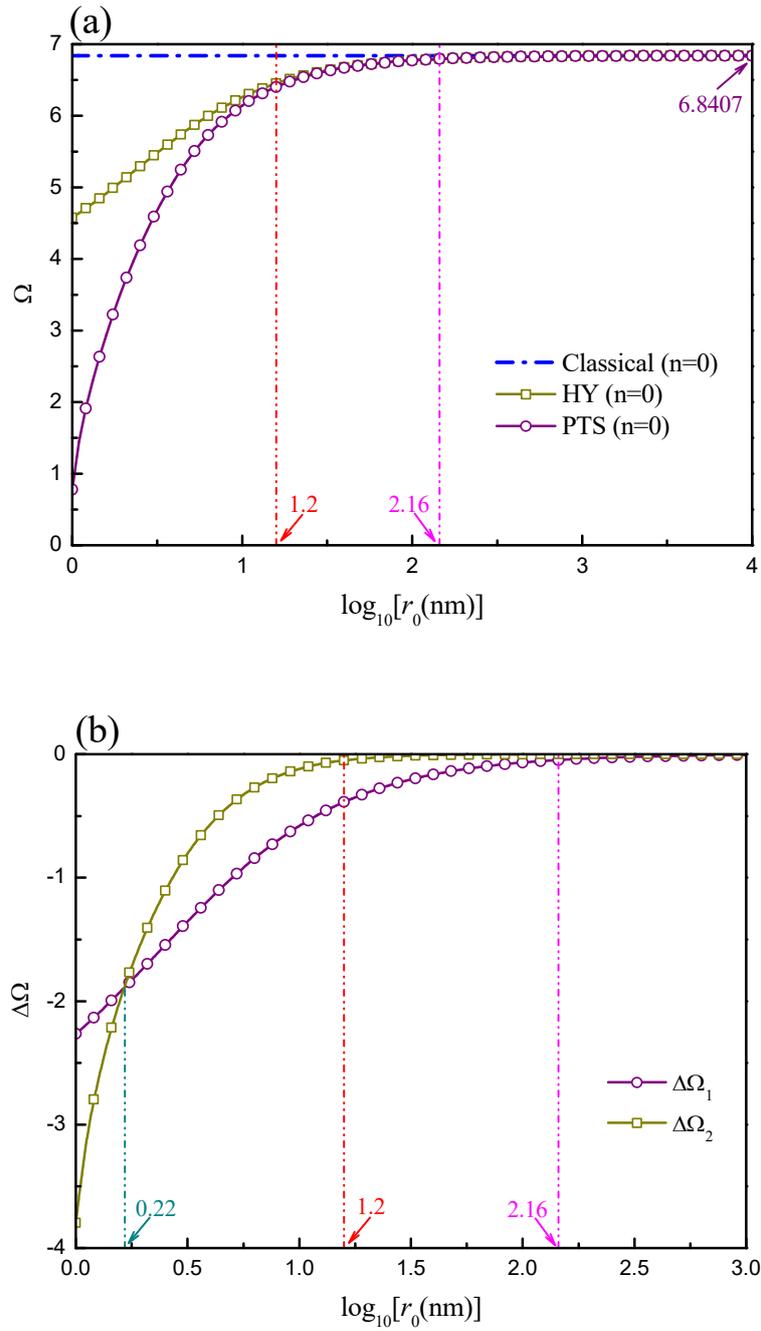

**Figure 4.** (a) Non-dimensional frequencies for the breathing mode $n = 0$ of the second class versus $\log_{10}[r_0(\text{nm})]$ for both HY and PTS theories; (b) Non-dimensional frequency differences $\Delta\Omega_1$ and $\Delta\Omega_2$ versus $\log_{10}[r_0(\text{nm})]$ for the breathing mode.

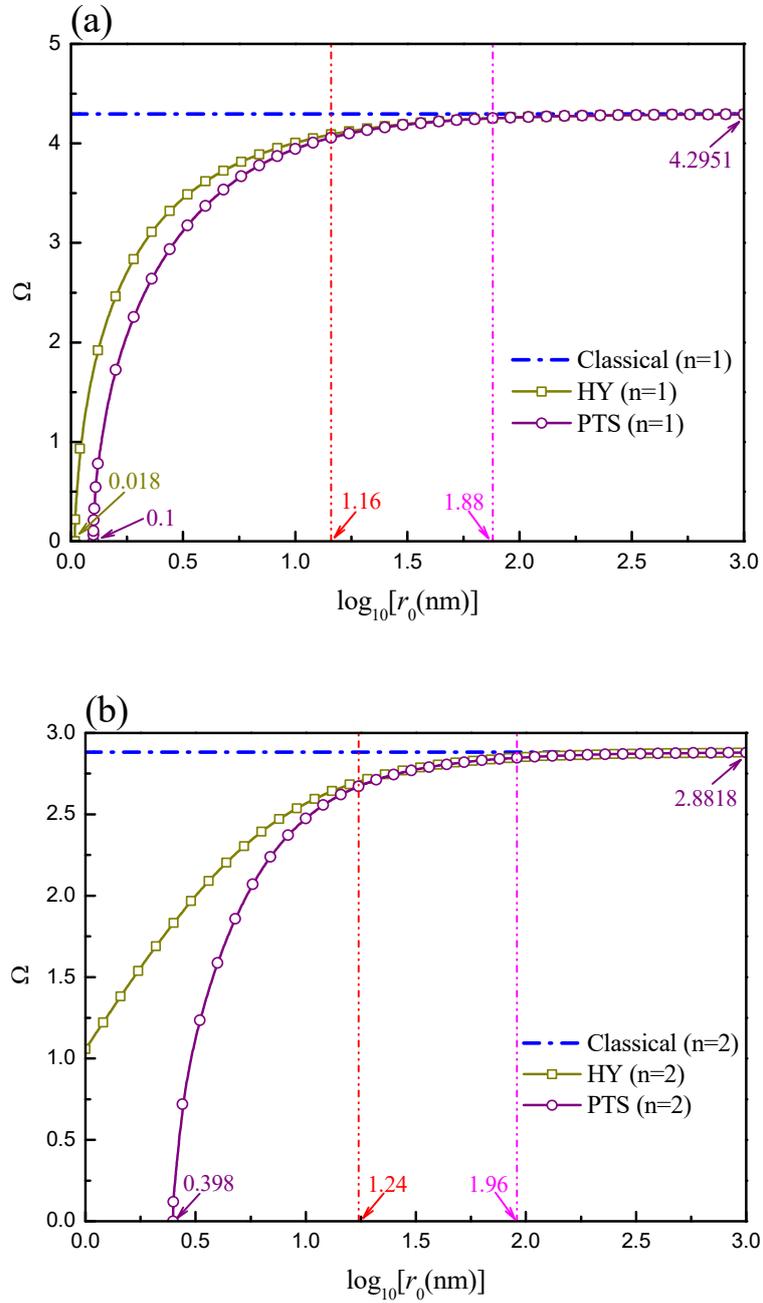

**Figure 5.** Non-dimensional frequencies for the non-breathing modes of the second class versus $\log_{10}[r_0(\text{nm})]$ for both HY and PTS theories: (a) $n=1$; (b) $n=2$.

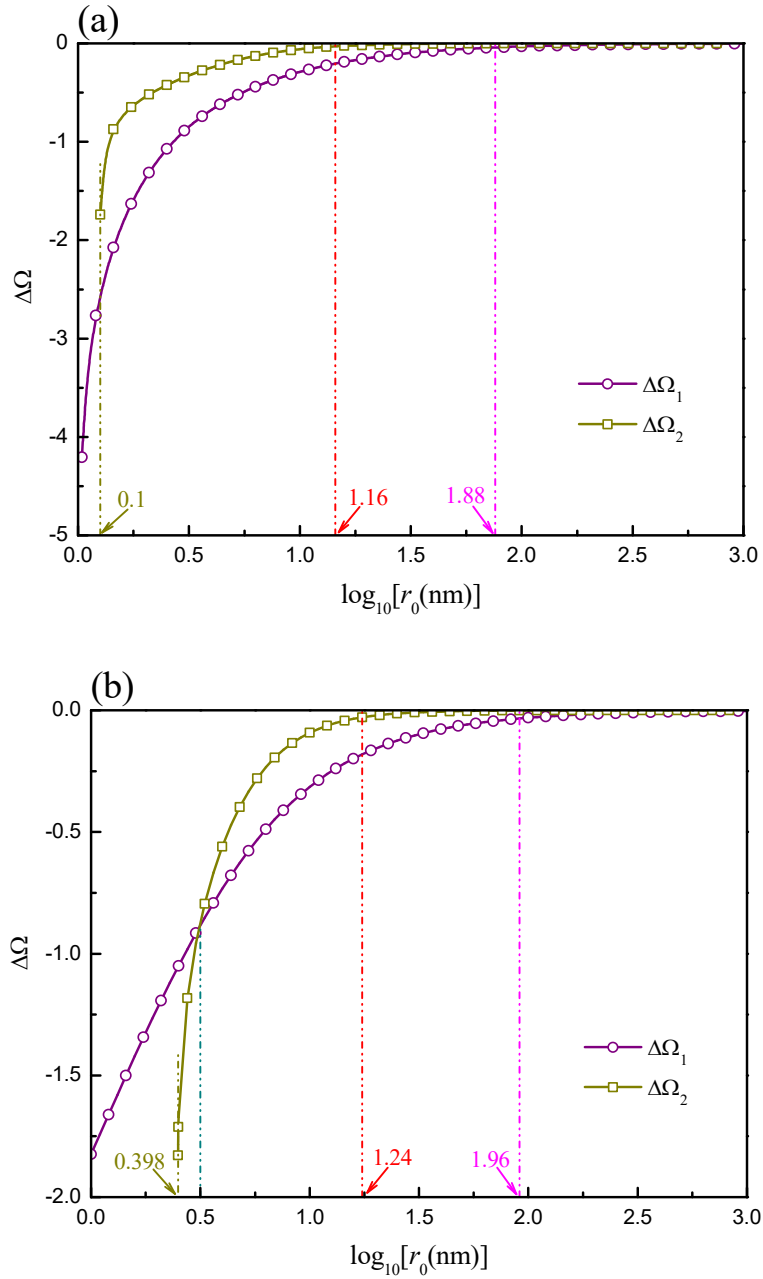

**Figure 6.** Non-dimensional frequency differences $\Delta\Omega_1$ and $\Delta\Omega_2$ versus $\log_{10}[r_0(\text{nm})]$ for the non-breathing modes of the second class: (a) $n=1$; (b) $n=2$.